\documentclass{PoS}

\title{Staggered fermions simulations on GPUs}

\ShortTitle{Staggered fermions on GPUs}

\author{\speaker{Claudio Bonati}\\
        Dipartimento di Fisica, Universit\`a di Pisa and INFN, Largo Pontecorvo 3, I-56127 Pisa, Italy.\\
        E-mail: \email{claudio.bonati@pi.infn.it}}

\author{Guido Cossu\\
        KEK Theory Center, 1-1 Oho, Tsukuba-shi, Ibaraki 305-0801, Japan. \\
        E-mail: \email{cossu@post.kek.jp}}

\author{Massimo D'Elia\\
        Dipartimento di Fisica, Universit\`a di Genova and INFN, Via Dodecaneso 33, 16146 Genova, Italy.\\
        E-mail: \email{massimo.delia@ge.infn.it}}

\author{Adriano Di Giacomo\\
        Dipartimento di Fisica, Universit\`a di Pisa and INFN, Largo Pontecorvo 3, I-56127 Pisa, Italy.\\
        E-mail: \email{digiaco@df.unipi.it}}

\abstract{We present our implementation of the RHMC algorithm for staggered fermions on Graphics Processing Units using the NVIDIA CUDA programming 
language. While previous studies exclusively deal with the Dirac matrix inversion problem, our code performs the complete MD trajectory on the GPU. 
After pointing out the main bottlenecks and how to circumvent them, we discuss the performance of our code.}

\FullConference{The XXVIII International Symposium on Lattice Field Theory\\
		 June 14-19,2010\\
		 Villasimius, Sardinia Italy}

\def\eg{\emph{e.g.} }
\def\ie{\emph{i.e.} }

\begin{document}

\section{Introduction}

In recent years the video game market developments compelled graphic processing units (GPUs) manufacturers to increase the floating point calculation performance of their products, by far exceeding the performance of standard CPUs. The architecture evolved toward programmable many-core chips that are designed to process in parallel massive amounts of data.  These developments suggested the possibility of using GPUs in the field of high-performance computing as low-cost substitutes of more traditional CPU-based architectures. 

The introduction of GPUs in lattice QCD calculations started with the seminal work of Ref.~\cite{videogame}, in which the native graphics APIs were used, but the real explosion of interest in the field followed the introduction of NVIDIA's CUDA (Compute 
Unified Device Architecture) platform, that effectively disclosed the field of GPGPU (General Purpose GPU \cite{cuda}). 

Previous studies on the application of GPUs to lattice QCD calculations were mainly aimed at using them together with the standard architectures in order to speed up some 
specific steps, typically the expensive Dirac matrix inversion. 
Our intent is to use GPUs in substitution of the usual architectures, actually performing the whole 
simulation by them. To achieve this result we found simpler to write a complete program from scratch instead of using 
existing software packages\footnote{On earlier stage we wrote a staggered version of JLab's  Chroma working on GPUs.}, in order to have a better control of all the steps to be performed and ultimately transferred 
to the GPU. Our implementation uses NVIDIA's CUDA platform together with a standard C++ serial control program running on CPU.

\section{The algorithm}

To simulate $N_f$ flavours of staggered fermions the Rational Hybrid Monte Carlo (RHMC) algorithm, introduced in 
\cite{rhmc}, has become the standard choice. We used standard (\ie non-improved) staggered fermions and, to speed-up the 
simulations, the following common tricks were implemented
\begin{itemize}
\item even/odd preconditioning
\item multi-step integrator (action divided in gauge and fermion part)
\item improved integrator (second order minimum norm)
\item multiple pseudo-fermions to reduce the fermion force magnitude and increase integration step size
\item different rational approximations and stopping residuals for Metropolis step and Molecular Dynamic (MD)
\end{itemize}

\section{A GPU scratch}

\begin{figure}
\centering
\scalebox{0.2}{\rotatebox{0}{\includegraphics{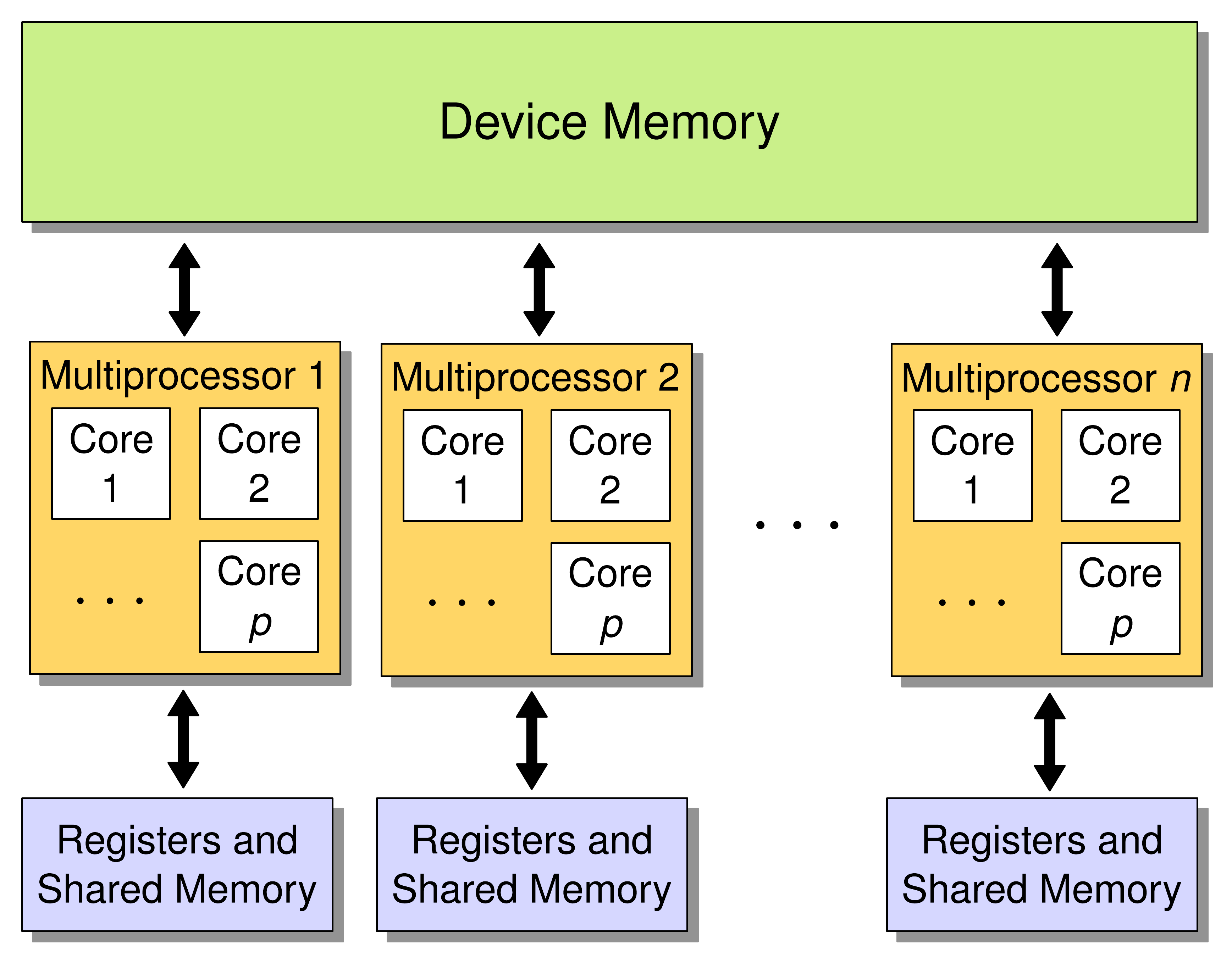}}}
\caption{Architecture of a modern NVIDIA graphics card.}
\label{arch_fig}
\end{figure}

NVIDIA GPUs are massively parallel computing elements, composed of hundreds of cores (called streaming processors) grouped
into multiprocessors. The typical architecture of a modern NVIDIA graphic card is outlined in Fig.~\ref{arch_fig}.

Three different storage levels are present: primary storage is provided by the device memory, which is accessible by all multiprocessors but has a relatively high latency. Within the same multiprocessor, cores have access to local registers and to shared memory, which is shared between the threads of the multiprocessor and it is orders of magnitude faster than device memory, being very close to the computing units. While the total amount of device memory is of order of GBs, the local storage is only 16KB both for the registers and for the shared memory\footnote{For NVIDIA Tesla cards 10 series. The 20 series has 64KB of on-chip memory that can be partitioned as shared and L1 cache.}, so that it is typically impossible to use just these local fast memories. 
The latency time of the device memory can be hidden by having a large number of threads in concurrent execution, so when data are needed from device memory for some threads, the ones ready to execute are immediately sent to computation. The highest bandwidth from device memory is achieved when a group of 16 threads accesses a contiguous memory region (coalesced memory access), because its execution requires just one instruction call, saving a lot of clock-cycles. This will be crucial in the following, when discussing the storage model for the gauge configuration.

\begin{table}[b]
\centering
\begin{tabular}{|l|l|l|l|l|l|}
\hline 
GPU  & Cores & Bandwidth & Gflops (peak) & Gflops (peak) & Device Memory \\
     &       & GB/s      & single        & double        & GB            \\
\hline\hline
Tesla C1060 & 240 & 102 & 933 & 78 & 4 \\
\hline 
Tesla C2050/2070 & 448 & 144 & 1030 & 515 & 3/6 \\
\hline
\end{tabular}
\caption{Specifications of the NVIDIA cards used in this work.}\label{spec_tab}
\end{table}

Double precision capability was introduced with NVIDIA's GT200 generation, the first one specifically designed having in mind HPC market, and
by now there is only a factor $2$ between the peak performance in single and double precision. In Tab.~\ref{spec_tab} the specifications of the GPUs used in this work are reported. 

Communications between the GPU and the CPU host are settled by a PCI express bus, whose typical bandwidth is 5GB/s, to be
compared with the GPU internal bandwidth between device memory and cores of order 100 GB/s. This is clearly the main bottleneck in most of GPU applications. 
We thus decided to copy the starting gauge configuration (and momenta) on the device memory at the beginning of the 
simulation and to perform the complete update on the GPU, instead of using it just to speed up some functions and transferring gauge field back and forth between host and device memories. 
  
In our implementation of the Dirac kernel a different thread is associated to every (even) site in the fermion update and to every link
in the gauge update, so that different treads do not cooperate. Shared memory is thus used just as a local fast memory. This setup is forced by the high ratio between data and floating point operations per kernel.

\section{Precision issues}

We will now address the issues related to the use of double precision. The main drawback of double precision is clear from Tab.~\ref{spec_tab}: single precision floating point arithmetic always outperforms the double one, although in the Fermi architecture the double precision penalty was significantly reduced. Another motivation to prefer the single precision is to speed up memory transfers because lattice QCD calculations are typically bandwidth limited.

Nevertheless double precision appeared to be necessary in the evaluation of the action for the Metropolis step to be performed at the end of a MD trajectory, which guarantees the correctness of the RHMC algorithm (see also Sec.~\ref{inv_sec}). Because of that the first and the last Dirac inversions are performed in double precision, while the inversions needed in the fermion force calculation are in single precision. The update of the gauge field is performed in single precision and double precision is used only in the reunitarization. 

Another important feature that is necessary for the algorithm to be exact is the reversibility of the trajectories \cite{hmc}. Since the gauge updates use only single precision we can expect reversibility to be valid only up to single precision and this is indeed the case: the magnitude of the reversibility violation for degree of freedom is measured to be of order $5\times 10^{-6}$.

\section{Memory allocation scheme}

\begin{figure}
\centering
\setlength{\unitlength}{1cm}
\begin{picture}(10,2.8)

\put(0,0){\line(1,0){10}}
\put(0,0.7){\line(1,0){10}} 
\put(0,1.4){\line(1,0){10}}
\put(0,2.1){\line(1,0){10}} 
\put(0,2.8){\line(1,0){10}}  

\put(0,0){\line(0,1){2.8}}   
\put(1,0){\line(0,1){2.8}}   
\put(2,0){\line(0,1){2.8}}   
\put(3,0){\line(0,1){2.8}}   
\put(4,0){\line(0,1){2.8}}   
\put(5,0){\line(0,1){2.8}}   
\put(6,0){\line(0,1){2.8}}   
\put(7,0){\line(0,1){2.8}}   
\put(8,0){\line(0,1){2.8}}   
\put(9,0){\line(0,1){2.8}}   
\put(10,0){\line(0,1){2.8}}  

\footnotesize
\put(0.1,2.35){$b_{11}(1)$}
\put(1.1,2.35){$b_{11}(2)$} 
\put(2.1,2.35){$b_{11}(3)$} 
\put(3.3,2.35){$\cdots$} 
\put(4.3,2.35){$\cdots$} 
\put(5.1,2.35){$b_{12}(1)$}
\put(6.1,2.35){$b_{12}(2)$} 
\put(7.1,2.35){$b_{12}(3)$} 
\put(8.3,2.35){$\cdots$} 
\put(9.3,2.35){$\cdots$} 

\put(0.3,1.65){$\cdots$} 
\put(1.1,1.65){$b_{22}(1)$}
\put(2.1,1.65){$b_{22}(2)$} 
\put(3.1,1.65){$b_{22}(3)$} 
\put(4.3,1.65){$\cdots$} 
\put(5.3,1.65){$\cdots$} 
\put(6.1,1.65){$b_{23}(1)$}
\put(7.1,1.65){$b_{23}(2)$} 
\put(8.1,1.65){$b_{23}(3)$}
\put(9.3,1.65){$\cdots$} 

\put(0.1,0.95){$c_{11}(1)$}
\put(1.1,0.95){$c_{11}(2)$} 
\put(2.1,0.95){$c_{11}(3)$} 
\put(3.3,0.95){$\cdots$} 
\put(4.3,0.95){$\cdots$} 
\put(5.1,0.95){$c_{12}(1)$}
\put(6.1,0.95){$c_{12}(2)$} 
\put(7.1,0.95){$c_{12}(3)$} 
\put(8.3,0.95){$\cdots$} 
\put(9.3,0.95){$\cdots$} 

\put(0.3,0.25){$\cdots$} 
\put(1.1,0.25){$c_{22}(1)$}
\put(2.1,0.25){$c_{22}(2)$} 
\put(3.1,0.25){$c_{22}(3)$} 
\put(4.3,0.25){$\cdots$} 
\put(5.3,0.25){$\cdots$} 
\put(6.1,0.25){$c_{23}(1)$}
\put(7.1,0.25){$c_{23}(2)$} 
\put(8.1,0.25){$c_{23}(3)$}
\put(9.3,0.25){$\cdots$} 
\end{picture}
\caption{Gauge field storage model: the element $u_{ij}(k)$ of the link $k$ is given by $u_{ij}(k)=b_{ij}(k)+c_{ij}(k)$.
$b_{ij}$ and $c_{ij}$ are respectively the most and the less significant bits of $u_{ij}$.} 
\label{gauge_fig}
\end{figure}

We noted previously that a correct allocation scheme is of the utmost importance in order to efficiently use the device memory. For staggered fermions, the storage of the gauge configuration is the most expensive one, so we will concentrate
on this. Similar techniques can be used also for the momenta and the pseudo-fermions storage. 

As stated before, QCD calculations on GPU are typically bandwidth limited and so it is convenient not to storage all the elements of an $SU(3)$ matrix (18 real numbers), but to use a representation in terms of fewer parameters. 
In this way we can reduce the amount of memory exchange at the expense of increasing the computational complexity. The additional calculations do not introduce significant overhead, actually they are negligible compared to the memory transfers. We used a 12 real number representation: only the first two rows are stored, while the third is reconstructed on fly.

Since in the Metropolis step the inversion of the Dirac matrix in double precision is required, we need to store a double precision gauge configuration, although in most of the calculations it will be used just as a single precision one.
In order not to waste bandwidth and device memory, it is useful to write a double precision number $a$ by using two
single precision numbers $b$ and $c$: $b$ is defined by the 32 most significant bits of $a$, while $c$ stores the
32 less significant ones. In C language this amounts to
\begin{eqnarray*}
&& b=\mathtt{(float)} a\\
&& c=\mathtt{(float)} (a-\mathtt{(double)}b)
\end{eqnarray*}
When only single precision is required we can just use $b$ instead of $a$, otherwise we have two possible choices:
to use $b$ and $c$ directly, effectively avoiding the explicit use of double precision arithmetic (see \eg \cite{gst}),
or to reconstruct the double precision number $a$ to be used in calculations. We implemented this last method, which is expected to be more efficient on double precision capable hardware. 

\begin{table}[t]
\centering
\begin{tabular}{|l|l|l|}
\hline
Lattice & Bandwidth GB/s & Gflops \\
\hline
$4\times 16^3$ & $56.84\pm 0.03$ & $49.31 \pm 0.02$ \\
\hline
$32\times 32^3$ & $64.091\pm 0.002$ & $55.597\pm 0.002$ \\
\hline
$4\times 48^3$ & $69.94\pm 0.02$ & $60.67\pm 0.02$ \\
\hline
\end{tabular}
\caption{Staggered Dirac operator kernel performance figures on a C1060 card (single precision).}\label{perf_tab}
\end{table}

To get coalesced memory accesses it is crucial for blocks of thread in execution to use contiguous regions of device memory.
This behaviour is maximized if we adopt the storage model shown in Fig.~\ref{gauge_fig};  the use of texture memory is a further improvement to reduce the effects of imperfect memory accesses.  

The performance of the Dirac operator in single precision which is obtained by means of this storage scheme is shown in Tab.~\ref{perf_tab}. From these data it is clear that the main bottleneck is the bandwidth: while using $60-70\%$ of the bandwidth, only the $5-6\%$ of the peak performance is reached.

\section{Inverter}\label{inv_sec}

The inversion of the Dirac operator in lattice QCD simulations is usually performed by using Krylov space solvers; for
staggered fermions the optimal choice is the simplest one of this class of solvers: the Conjugate Gradient (CG) algorithm.

\begin{figure}[b]
\centering
\includegraphics[width=0.48\textwidth,clip]{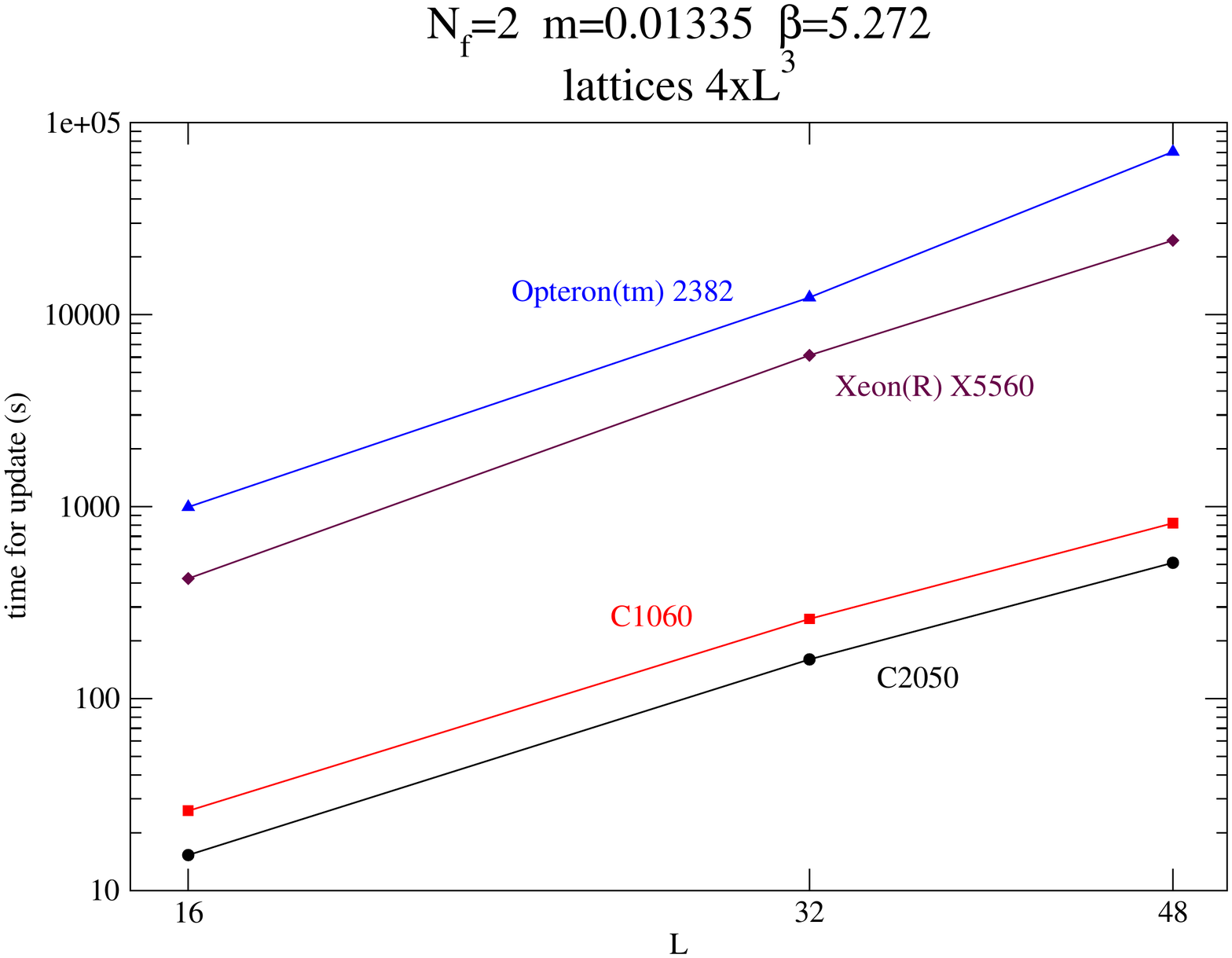}
\includegraphics[width=0.48\textwidth,clip]{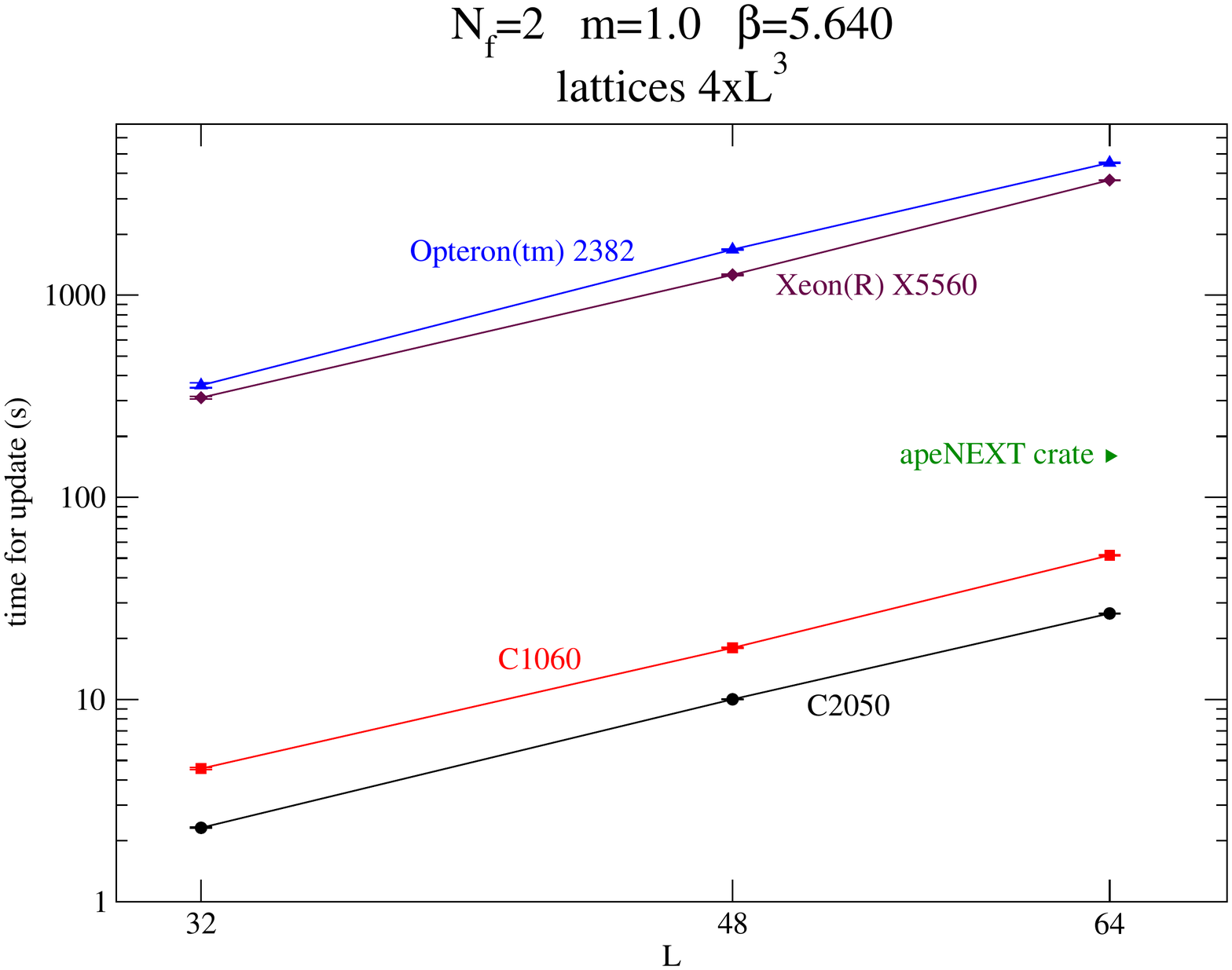}
\caption{Run times on different architectures. For the Opteron and Xeon runs a single core was used.}
\label{time_fig}
\end{figure}

In all Krylov space solvers the approximate solution and its estimated residual are calculated recursively. While in exact arithmetic the estimated residual is exactly the residual of the approximate solution, in finite precision this is no more the case. The attainable accuracy at fixed precision can be estimated \cite{greenbaum} and for a single precision Dirac inversion a typical value for the minimum true residual is $10^{-2}-10^{-3}$, so we need double precision to perform the inversions related to the Metropolis step.

In standard Krylov solvers this problem can be overcome by using the residual replacement strategy: sometimes the true residual is explicitly calculated in double precision and the algorithm is restarted. With this method it is possible to obtain reliable results, as with double precision calculations, but using almost always single precision arithmetic. Residual replacement methods are well understood theoretically \cite{vy} and have been successfully applied to QCD calculations on GPUs \cite{clark}. However, in RHMC we need Krylov solvers for shifted systems (see \eg \cite{shift}), whose starting solution has to be the null one, thus preventing the restarting of the algorithm. For this reason the Dirac inversions in the Metropolis step have to be performed completely in double precision.

\section{Performance}

In Fig.~\ref{time_fig} the RHMC update time on different architectures is shown for two values of the bare quark masses ($am=0.01335, 1.0$). For both the mass values the scaling with the size of the lattice is good. In fact it is a characteristic feature of GPUs that increasing the lattice size improves the computational efficiency, as seen also in Tab.~\ref{perf_tab}; this happens because with large lattices internal latencies are hidden more effectively.  Time gains for Tesla C1060 and C2050 are shown in Tab.~\ref{1060_tab} and Tab.~\ref{2050_tab}; particularly impressive is the comparison with the results obtained by using an apeNEXT crate.

\begin{table}
\centering
\begin{tabular}{|l|l|l|l|l|l|l|}
\hline
{ } &  \multicolumn{3}{c}{high mass} \vline & \multicolumn{3}{c}{low mass} \vline \\ 
\hline
spatial size  &  32 & 48 & 64 & 16 & 32 & 48 \\ \hline\hline
Opteron (single core) & 65 & 75 & 75 & 40 & 50 & 85 \\ \hline
Xeon (single core) & 50 & 50 & 50 &  15 & 25 & 30 \\ \hline
apeNEXT crate   & \multicolumn{3}{c}{\(\sim\)3}\vline & \multicolumn{3}{c}{\(\sim\)1}\vline \\ \hline
\end{tabular}
\caption{NVIDIA C1060 time gains over CPU and apeNEXT.} \label{1060_tab}
\end{table} 

\begin{table}
\centering
\begin{tabular}{|l|l|l|l|l|l|l|}
\hline
{ } &  \multicolumn{3}{c}{high mass} \vline & \multicolumn{3}{c}{low mass} \vline \\ 
\hline
spatial size  &  32 & 48 & 64 & 16 & 32 & 48 \\ \hline\hline
Opteron (single core) & 115 & 130 & 140 & 65 & 75 & 140\\ \hline
Xeon (single core) & 85 & 85 & 100 &  30 & 40 & 50 \\ \hline
apeNEXT crate   & \multicolumn{3}{c}{\(\sim\)6}\vline & \multicolumn{3}{c}{\(\sim\)2}\vline \\ \hline
\end{tabular}
\caption{NVIDIA C2050 time gains over CPU and apeNEXT (same code as for C1060, no specific C2050 improvement implemented).}\label{2050_tab}
\end{table}

\section{Conclusions}

The extremely high computation capabilities of modern GPUs make them attractive platforms for high-performance computations. Previous studies on lattice QCD applications have been devoted almost exclusively to the Dirac matrix inversion problem. We have shown that it is possible to use GPUs to efficiently perform a complete simulation, without the need to rely on more traditional architectures.

\section{Acknowledgment}
It's a pleasure to thank Edmondo Orlotti (NVIDIA) and Massimo Bernaschi (IAC, Rome) for the possibility they gave us to test our code on a C2050 card.


\begin{thebibliography}{99}
 
\bibitem{videogame} G.~I.~Egri, Z.~Fodor, C.~Hoelbling, S.~D.~Katz, D.~Nogradi, K.~K.~Szabo 
Comput. Phys. Commun. {\bf 177}, 631 (2007) [{\tt arXiv:hep-lat/0611022}].

\bibitem{cuda} NVIDIA Corporation, http://developer.nvidia.com/object/gpucomputing.html

\bibitem{rhmc} I.~Horvath, A.~D.~Kennedy, S.~Sint 
Nucl. Phys. B Proc. Suppl. {\bf 73}, 834 (1999) [{\tt arXiv:hep-lat/9809092}].

\bibitem{hmc} S.~Duane, A.~D.~Kennedy, B.~J.~Pendleton, D.~Roweth 
Phys. Lett. B {\bf 195}, 216 (1987). 

\bibitem{gst}  D.~G\"{o}ddeke, R.~Strzodka, S.~Turek 
International Journal of Parallel, Emergent and Distributed Systems  {\bf 22}, 221 (2007).

\bibitem{greenbaum} A.~Greenbaum 
SIAM J. Matrix Anal. Appl. {\bf 18}, 535 (1997).

\bibitem{vy} H.~K.~van der Vorst, Q.~Ye 
SIAM J. Sci. Comput. {\bf 22}, 835 (2000).

\bibitem{clark} M.~A.~Clark, R.~Babich, K.~Barros, R.~C.~Brower, C.~Rebbi 
Comput. Phys. Commun. {\bf 181}, 1517 (2010) 
[{\tt arXiv:0911.3191} [hep-lat]].

\bibitem{shift} B.~Jegerlehner
[{\tt arXiv:hep-lat/9612014}]. 

\end{thebibliography}
\end{document}